\documentclass[%
preprint, 
amsmath,amssymb,
aps, physrev,
]{revtex4-2}
\DeclareUnicodeCharacter{2212}{-}
\usepackage{graphicx}
\usepackage{dcolumn}
\usepackage{bm}
\usepackage{multirow}

\usepackage{verbatim}
\usepackage{mathrsfs}
\usepackage{amsmath,amsfonts,amssymb}
\usepackage{bbm}
\usepackage{graphics}
\usepackage{amsmath}
\let\Gamma\varGamma
\let\Delta\varDelta
\let\Theta\varTheta
\let\Lambda\varLambda
\let\Xi\varXi
\let\Pi\varPi
\let\Sigma\varSigma
\let\Upsilon\varUpsilon
\let\Phi\varPhi
\let\Psi\varPsi
\let\Omega\varOmega

\begin{document}
	\preprint{}
	\title{\textbf{Existence of dual topological phases in \textit{Sn}-based ternary chalcogenides}}%
	\author{Ramesh Kumar}
	\author{Mukhtiyar Singh}%
	\email{mukhtiyarsingh@dtu.ac.in; msphysik09@gmail.com}
	\affiliation{%
		\\
		Computational Quantum Material Design (CQMD) Lab, Department of Applied Physics, Delhi Technological University, New Delhi-110042, India}%
		
	\begin{abstract}
		It is quite intriguing to investigate the transition from a topological insulator (TI) phase to topological crystalline insulator (TCI) phase in a material as the latter has an advantage over the former in controlled device applications. This work investigates the existence of this dual topological behavior in \textit{Sn}-based ternary chalcogenides family PbSnX$_2$ (X=S, Se, Te) under the hydrostatic pressure using first-principles approach. These materials are dynamically stable at ambient and elevated pressure conditions up to which the topological phase transitions (TPTs) are studied.  This family have a topologically trivial ground state with direct band gap values 0.338 eV, 0.183 eV and 0.217 eV for PbSnS$_2$, PbSnSe$_2$ and {PbSnTe$_2$, respectively. The first TPT i.e., TI phase for these materials is observed, under the effect of external pressure of 5 GPa, 2.5 GPa and 3.5 GPa, with a single band inversion at\textit{ F-point} in the bulk band structure and an odd number of Dirac cones along the (111) surface. A further increase in pressure to 5.5 GPa, 3 GPa and 4 GPa results in another band inversion at \textit{$\Gamma$-point} and an even number of Dirac cones along the (111) plane. These even number of band inversions suggest that (\(\bar{1}2\bar{1}\)) surface has mirror symmetry around (\(\bar{1}0\bar{1}\)) plane and hence, the TCI phase is obtained. This TCI phase is further corroborated with even value of mirror Chern number calculated using winding of Wannier charge centers. 
		\begin{description}
			\item[Keywords]
			Topological phase transition, ternary chalcogenides, first-principles calculations, topological crystalline insulators, mirror Chern number.
		\end{description}
}
	\end{abstract}
	\maketitle
	\section{\label{sec:level1}INTRODUCTION\protect}
	A new phase of matter with unusual metallic edge/surface states known as topological materials (TM) has been studied intensively in recent years \cite{1,2,3,4,5,6}. The TMs can be divided into few categories such as topological insulators (TIs) \cite{3, 4, 5, 6}, topological crystalline insulators (TCIs) \cite{7, 8, 9}, Dirac semimetals (DSMs) \cite{10, 11}, Weyl semimetals (WSMs) \cite{12, 13}, nodal line semimetals (NLSMs) \cite{14, 15}, $\mathrm{Z_2}$ topological semimetals \cite{16, 17} and triply degenerate node semimetals \cite{18, 19}, etc. The metallic Dirac-like electronic states on the surface of the crystal with insulating bulk are the signature of the existence of TI. These metallic edge/surface states appear in the presence of spin-orbit coupling (SOC) which are protected by time-reversal symmetry (TRS) and provides robust spin-polarized conduction channels \cite{1, 2, 3, 4, 5, 6}. These materials have potential applications in low-power high-speed electronic and spintronic devices, quantum computing, and thermoelectric applications \cite{20, 21, 22} etc. The careful alteration in the strength of SOC plays an important role in realizing topological phases in specific materials. The hydrostatic pressure is a non-destructive way of enhancing SOC without affecting the charge neutrality of the materials. The binary and ternary materials like $\mathrm{Bi_{4}Br_{4}}$ \cite{23}, LaAs \cite{16}, YbAs \cite{24}, LnSb (Ln=La, Gd, Tm) \cite{17, 25}, YX (X=As, Sb, Bi) \cite{26, 27} and $\mathrm{KNa_2Bi}$ \cite{28}, BiTeI \cite{29}, $\mathrm{Bi_2S_3}$ \cite{30}, BiTeBr \cite{31} have shown the topological phase transition (TPT) with hydrostatic pressure.
	
	The term “topological crystalline insulators,” has been introduced by L. Fu, which defines a class of materials having robust topological states and are protected by crystalline symmetries \cite{7}. This new phase was first proposed in SnTe and later angle-resolved photoemission spectroscopy (ARPES) verified the topological surface states (TSSs) in the (001) plane of this material \cite{8}. Unlike TIs, TCIs have persisted protection by mirror symmetry even when TRS is broken. It is quite intriguing to investigate the TPT from TI to TCI in a material. TCI phase is more robust against impurities or defects and has diverse and tunable topological characteristic states \cite{7, 32}. TCI materials hold higher-order topological hinge or corner states which can be identified with different topological invariants depending on the crystal symmetries \cite{33, 34} and can have potential applications in stable qubits and novel electronic devices \cite{7, 32, 33, 34}. The TCI phase has an advantage over the TI phase for controlled device applications such as a topological field effect transistor \cite{32}. 
	
	The narrow band-gap semiconductors are amongst the highly sought-after materials to realize TPT. Several narrow gap IV-VI semiconducting materials such as PbTe, SnSe, SnS, TlSe, TlS, and $\mathrm{Pb_{1−x}Sn_{x}X}$ (X = Se, Te) have been studied for the existence of TIs and TCIs phases \cite{35, 36, 37, 38, 39}. Other than these, the ternary chalcogenides $\mathrm{TlXY_2}$ (X = Sb, Bi; Y = Te, Se) have also been reported as strong TIs \cite{40, 41} but the TCI phase has not been explored in these materials except for $\mathrm{TlBiS_2}$ and $\mathrm{TlSbS_2}$ \cite{42}. The present study focuses on the realization of TI and TCI phases in ternary chalcogenides $\mathrm{PbSnX_2}$ (X= S, Se, Te) with the application of external hydrostatic pressure.  For this family, trivial to non-trivial phase are obtained with single band inversion at $\Gamma$\textit{-point} under hydrostatic pressure. With further increases in pressure, a pair of Dirac cones are observed along the (111) plane and made these systems trivial again. However, a TCI phase emerges at this elevated pressure as confirmed by non-trivial TSSs along (\(\bar{1}2\bar{1}\)) plane which is symmetric around the mirror plane (\(\bar{1}0\bar{1}\)).  
	
	\section{\label{sec:level2}COMPUTATIONAL METHODOLOGY\protect}
	The structural optimization and electronic structure calculations based on the density functional theory (DFT) approach \cite{43, 44} were performed using the projector augmented-wave (PAW) \cite{45} method as implemented in the Vienna Ab initio Simulation Package (VASP) \cite{46}. The PAW potentials for Pb $\mathrm{(6s^26p^2)}$ with 4 valence electrons, Sn $\mathrm{(5s^25p^2)}$ with 4 valence electrons, and S $\mathrm{(3s^23p^4)}$, Se $\mathrm{(4s^24p^4)}$, Te $\mathrm{(5s^25p^4)}$ with 6 valance electrons configuration were used for the calculations. The exchange and correlation energy were calculated with the generalized gradient approximation of Perdew–Burke–Ernzerhof (GGA-PBE) \cite{47} and modified Becke-Johnson (mBJ) \cite{48} functionals. The effect of SOC was included in all calculations except in ionic optimization. The total energies convergence criterion of $\mathrm{10^{−6}}$ eV was adopted along with a finer 9x9x5 gamma-centered k-mesh. The plane-wave cut-off energies for $\mathrm{PbSnX_2}$ (X=S, Se, Te) were kept at 380 eV, 310 eV and 260 eV, respectively.
	
	The phonon calculations were performed using the PHONOPY code \cite{49}. The $\mathrm{Z_2}$ topological invariants were calculated using the product of parities at time reversal invariant momenta (TRIM) points as per the Kane and Mele model \cite{6}. Wannier90 code \cite{50} was used to obtain maximally localized Wannier functions (MLWFs) and to parametrize a tight-binding (TB) Hamiltonian. This TB Hamiltonian was used to obtain the surface density of state (SDOS) using the Green function approach as implemented in the WannierTools code \cite{51}.
	
	\section{\label{sec:level3}RESULTS AND DISCUSSION\protect}
	\subsection{\label{sec:level3}Crystal structure and stability analysis\protect}
	
	The ternary chalcogenides $\mathrm{PbSnX_2}$ (X=S, Se, Te) have adopted a rhombohedral crystal structure (space group \(R\bar{3}m\) (166)) (FIG. 1 (a)) and are isostructural to the well-studied $\mathrm{TlBiTe_2}$, $\mathrm{TlBiSe_2}$, and $\mathrm{TlBiS_2}$ materials \cite{52, 53, 54, 55, 56} which have also been experimentally synthesized (crystal structure in suppl. information (FIG. S1)). This family has a Pb-X-Sn-X-Pb- sequence along the \textit {three-fold axis}. The Pb/Sn layers are sandwiched between the chalcogenides layers as shown in FIG. 1 (a, b). The hexagonal supercell and the bulk Brillouin zone (BZ) with projected surfaces along (111) and (\(\bar{1}2\bar{1}\)) are shown in FIG. 1 (a, c). The surface BZ of the (111) and (\(\bar{1}2\bar{1}\)) planes are represented in FIG. S2. Both Pb and Sn atoms act as an inversion center with atomic coordinates Pb (0, 0, 0), Sn (0.5, 0.5, 0.5) and X (±v, ±v, ±v) sites. As hydrostatic pressure is isotropic in nature which, therefore, does not alter the symmetry of the systems for the entire studied pressure range. The optimized lattice parameters for this family are given in TABLE 1.
	\begin{table}[h]
		\centering
		\caption{Lattice parameters (in Å) of the ternary chalcogenides $\mathrm{PbSnX_2}$ (X=S, Se, Te) family. Also, the bulk band energy gap at $\Gamma$- and \textit{F-points} using TB-mBJ functional.}

		\begin{tabular}{|c|c|c|c|} \hline 
			Materials&  Lattice parameters&  Energy gap at \textit{F-point} (Bandgap, ${E_g}$)& Energy gap at \textit{$\Gamma$-point}\\ \hline 
			PbSnS$_2$&  a = b = 4.201; c = 20.449&  0.338 eV & 0.377 eV\\ \hline 
			PbSnSe$_2$&  a = b = 4.343; c = 21.262&  0.183 eV & 0.215 eV\\ \hline 
			PbSnTe$_2$&  a = b = 4.589; c = 22.538&  0.217 eV& 0.235 eV \\ \hline
		\end{tabular}
		\label{tab:my_label}
	\end{table}
	
	\begin{figure}[h]
		\centering
		\includegraphics[width=1\linewidth]{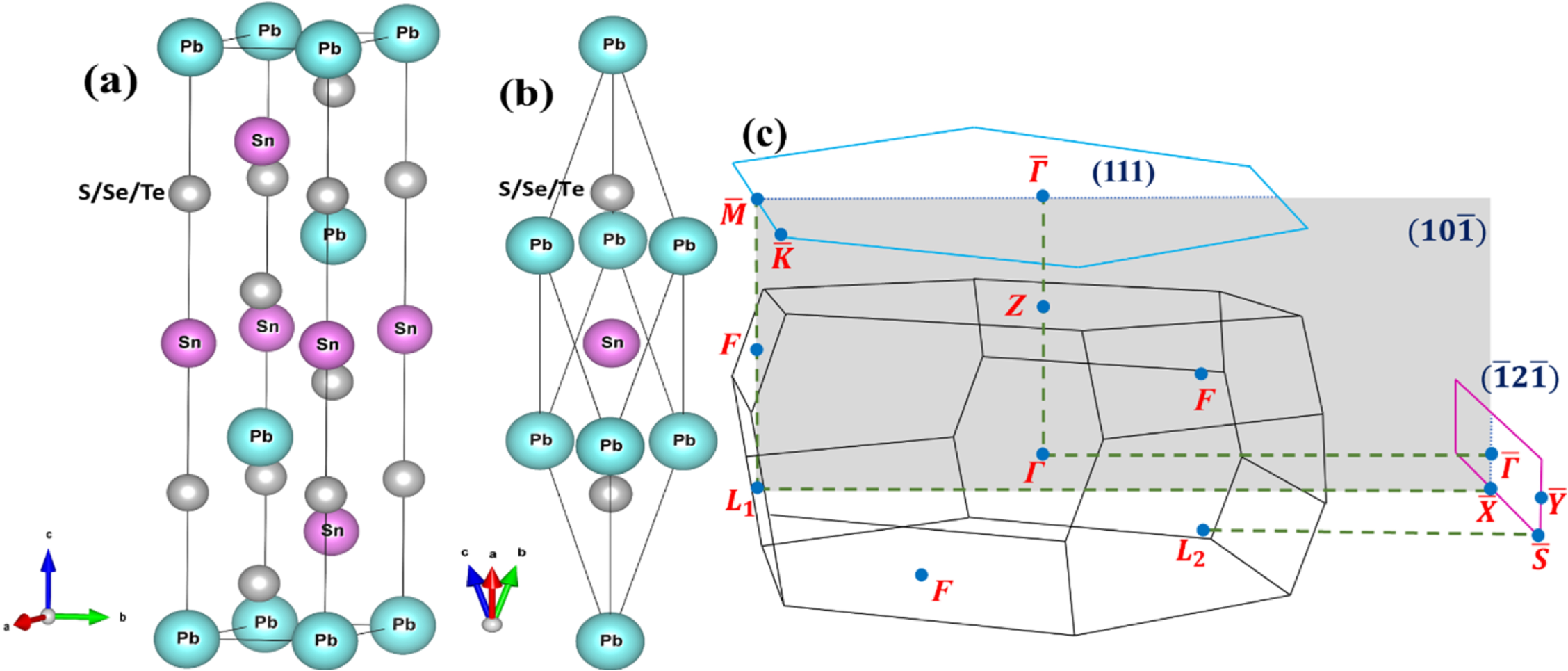}
		\caption{The crystal structure of $\mathrm{PbSnX_2}$ (X=S, Se, Te) (a) conventional hexagonal structure, (b) primitive unit cell, and (c) The bulk Brillion zone (BZ) of primitive rhombohedral cell with projections on surface Brillouin zones.}
		\label{fig:enter-label}
	\end{figure}
	
	The surface electronic structure for the TCI systems significantly depends on the orientation of the crystal face. We have considered two surfaces of the rhombohedral structure. ${L_1}$\textit{-} and \textit{F-points} are projected to \(\bar{M}\)\textit{-point} of the (111) plane whereas ${L_1-}$ and ${L_2-points}$ are projected on \(\bar{X}\) \textit{-} and \(\bar{S}\)\textit{-points}, respectively, on \(\bar{1}2\bar{1}\)  plane. The \(\bar{\Gamma}\)\textit{-points} as the center of both the planes are projections of the center of BZ i.e., $\Gamma$\textit{-point}. The (\(\bar{1}0\bar{1}\)) mirror plane is perpendicular to the (\(\bar{1}2\bar{1}\)) surface plane and projected along the  \(\bar{\Gamma}-\bar{X}\) direction as shown in FIG. 1 (c). Further, we have studied the stability of $\mathrm{PbSnX_2}$ (X=S, Se, Te) at ambient and elevated pressures via the phonon dispersion spectrum. The absence of negative frequencies at ambient and higher-pressures values as shown in FIG. S3 shows the stability of these materials. It has been noted that TPT in these materials takes place within this studied stability range of hydrostatic pressure.
	
	\subsection{\label{sec:level2}The electronic structure at ambient pressure}
	It is an established fact that the prediction of ground state using DFT is highly depend upon the choice of \textit{exchange-correlation} \textit{functional} corresponding to different material systems. To choose a relevant and accurate functional for our study, we performed the \textit{ab initio} calculation with two major functional i.e., GGA-PBE and mBJ for the experimentally synthesized $\mathrm{TlBiS_2}$ material, which is isostructural to our studied systems. We found that the mBJ functional predicts the true ground state of $\mathrm{TlBiS_2}$ with the band gap of 0.36 eV which is in excellent agreement with its experimental value of 0.35 eV \cite{52, 53, 54} (see suppl. FIG. S4). Therefore, we used mBJ functional for further investigations of $\mathrm{PbSnX_2}$ (X=S, Se, Te). However, the PBE-GGA has been quite successful in predicting the true ground state of binary chalcogenide materials \cite{35, 37, 39, 57, 58} but in the case of this ternary chalcogenide family, it predicts an inaccurate ground state (FIG. S5). FIG. 2 (a, c, e) depicts the bulk band structure of $\mathrm{PbSnX_2}$ (X=S, Se, Te) materials using mBJ functionals. The \textit{p-orbital} of the Pb/Sn in the conduction band (CB) and \textit{p-orbitals} of chalcogenide elements S/Se/Te in the valence band (VB) mainly contribute near the Fermi energy. 
	\begin{figure}[h]
		\centering
		\includegraphics[width=1\linewidth]{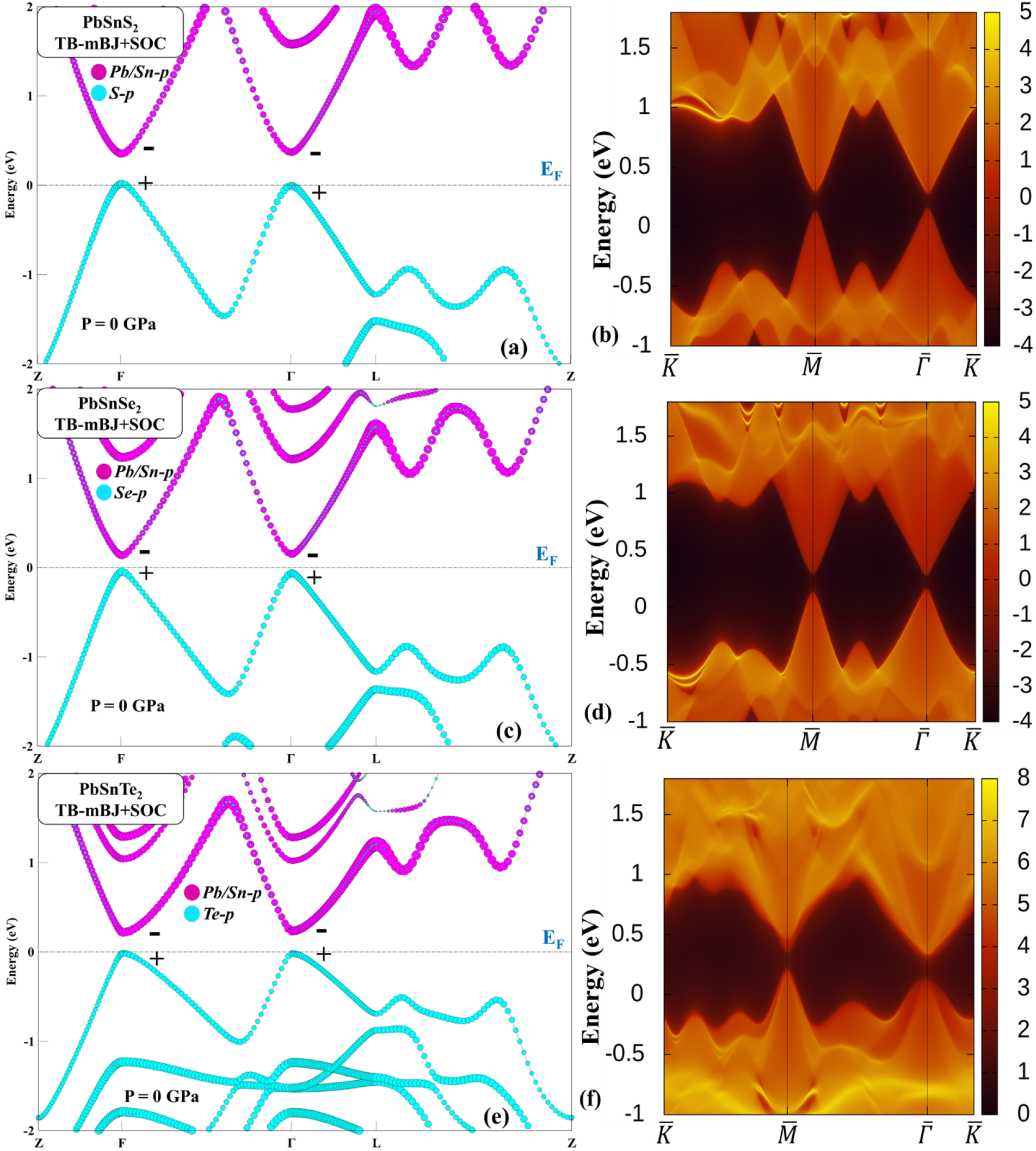}
		\caption{(a, c, e) The bulk band structure using mBJ+SOC functional, (b, d, f) The density of states (SODS) along (111) plane of the systems $\mathrm{PbSnX_2}$ (X=S, Se, Te) at ambient pressure.}
		\label{fig:enter-label}
	\end{figure}
	No inverted contribution of the orbitals near the Fermi level is observed in these materials and hence they are topologically trivial semiconductors with the direct band at $\Gamma$\textit{-point}. The energy gap of these materials at $\Gamma$\textit{-} and \textit{F-points} in bulk band structure is represented in TABLE 1. The absence of Dirac cones in SDOS along the (111) plane (FIG. 2 (b, d, f) in these materials also establishes their topologically trivial state at ambient pressure. 
	
	\subsection{\label{sec:level2}The effect of hydrostatic pressure on electronic structure}
	
	In the preceding section, we have established the topologically trivial semiconducting nature of $\mathrm{PbSnX_2}$ (X=S, Se, Te) at ambient pressure. Now, we have included the effect of hydrostatic pressure to tune the bulk electronic structure of these materials. The compressive nature of the hydrostatic pressure has performed the isotropic reduction in the lattice parameters and hence tuned the eigenstates of these systems. The application of hydrostatic pressure on $\mathrm{PbSnX_2}$ (X=S, Se, Te) results in the TPT (FIG. 3 (a, c, e)) at 5 GPa, 2.5 GPa and 3.5 GPa, respectively. An inverted contribution of \textit{p-orbitals} of chalcogenide elements S/Se/Te in CB and Pb/Sn in VB is observed at \textit{F-point}, whereas no change in orbital contribution is observed at $\Gamma$\textit{-point} as shown in the insets of FIG. 3 (a, c, e). The TPT pressure for $\mathrm{PbSnTe_2}$ is higher than $\mathrm{PbSnSe_2}$, which is due to the higher energy gap of this material at the \textit{F-point}. The SDOS along the (111) plane of these materials shows the presence of a single Dirac cone at $\Gamma$\textit{-point} in surface BZ as shown in FIG. 3 (b, d, f). 
	
	\begin{figure}
		\centering
		\includegraphics[width=1\linewidth]{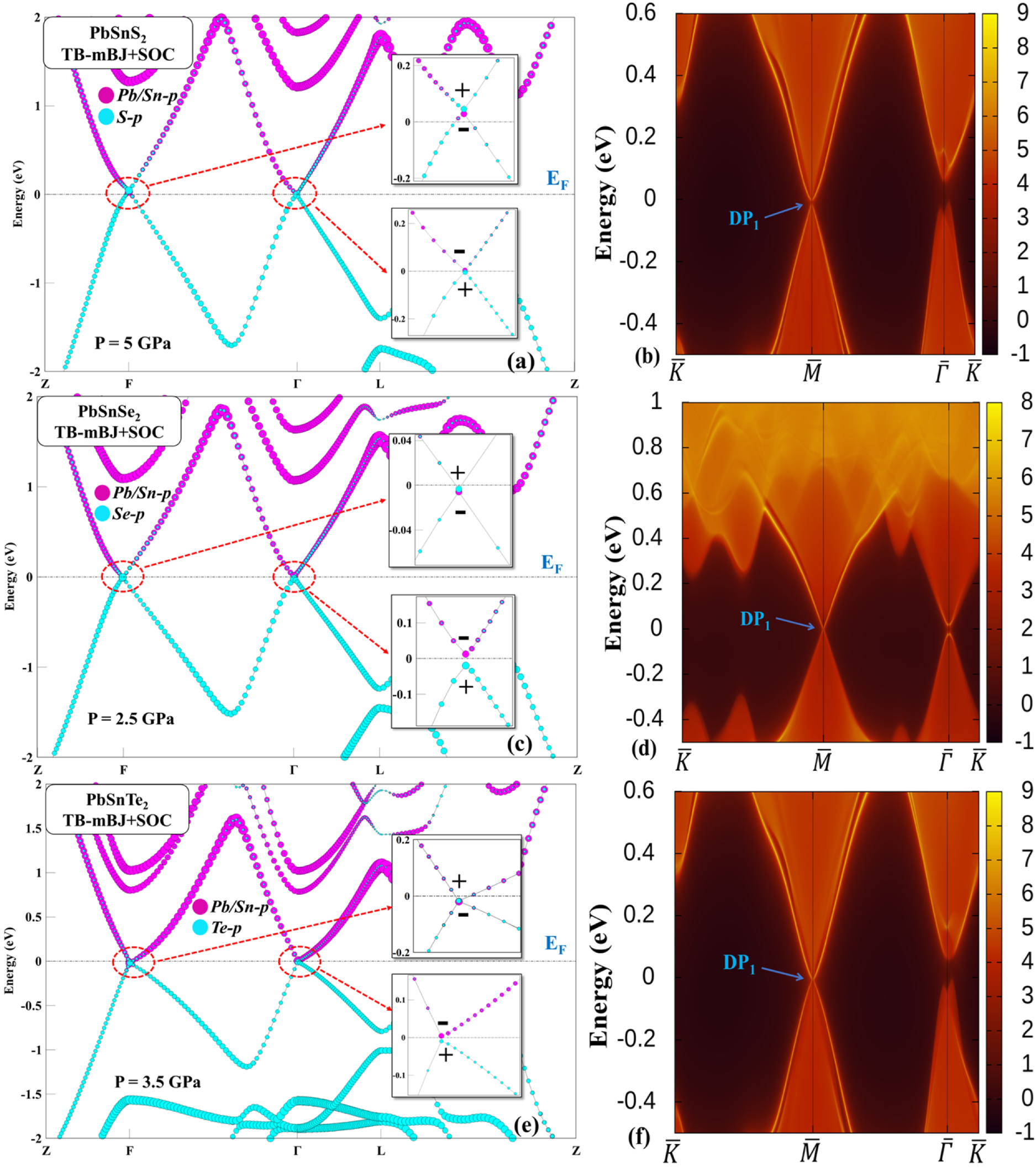}
		\caption{(a, c, e) The bulk band structure with band inversion at \textit{F-point}, (b, d, f) The surface density of states (SODS) along (111) plane of the systems $\mathrm{PbSnX_2}$ (X=S, Se, Te) at elevated pressure with marked Dirac point ($\mathrm{DP_1}$). }
		\label{fig:enter-label}
	\end{figure}
	
	A further increase in hydrostatic pressure tunes the bulk band structure at the $\Gamma$\textit{-point} and band inversion at this TRIM point also takes place as shown in FIG. 4 (a, c, e). The inverted contribution of the \textit{p-orbitals} of chalcogenide elements S/Se/Te in CB and Pb/Sn in VB has verified the TPT at $\Gamma$\textit{-point}. Now, there is an even number of band inversions in these systems which makes them topologically weak insulators or topologically trivial systems. A pair of Dirac cones in SDOS at \(\bar{\Gamma}\)\textit{-point} and \(\bar{M}\)\textit{-point} along the (111) plane as shown in FIG. 4 (b, d, f) also establishes the presence of an even number of band inversions in $\mathrm{PbSnX_2}$ (X=S, Se, Te). To verify whether these materials, with even number of band inversions, are topologically weak insulators or topologically trivial, further, we have calculated $\mathrm{Z_2}$ topological invariants.
	
	\begin{figure}
		\centering
		\includegraphics[width=1\linewidth]{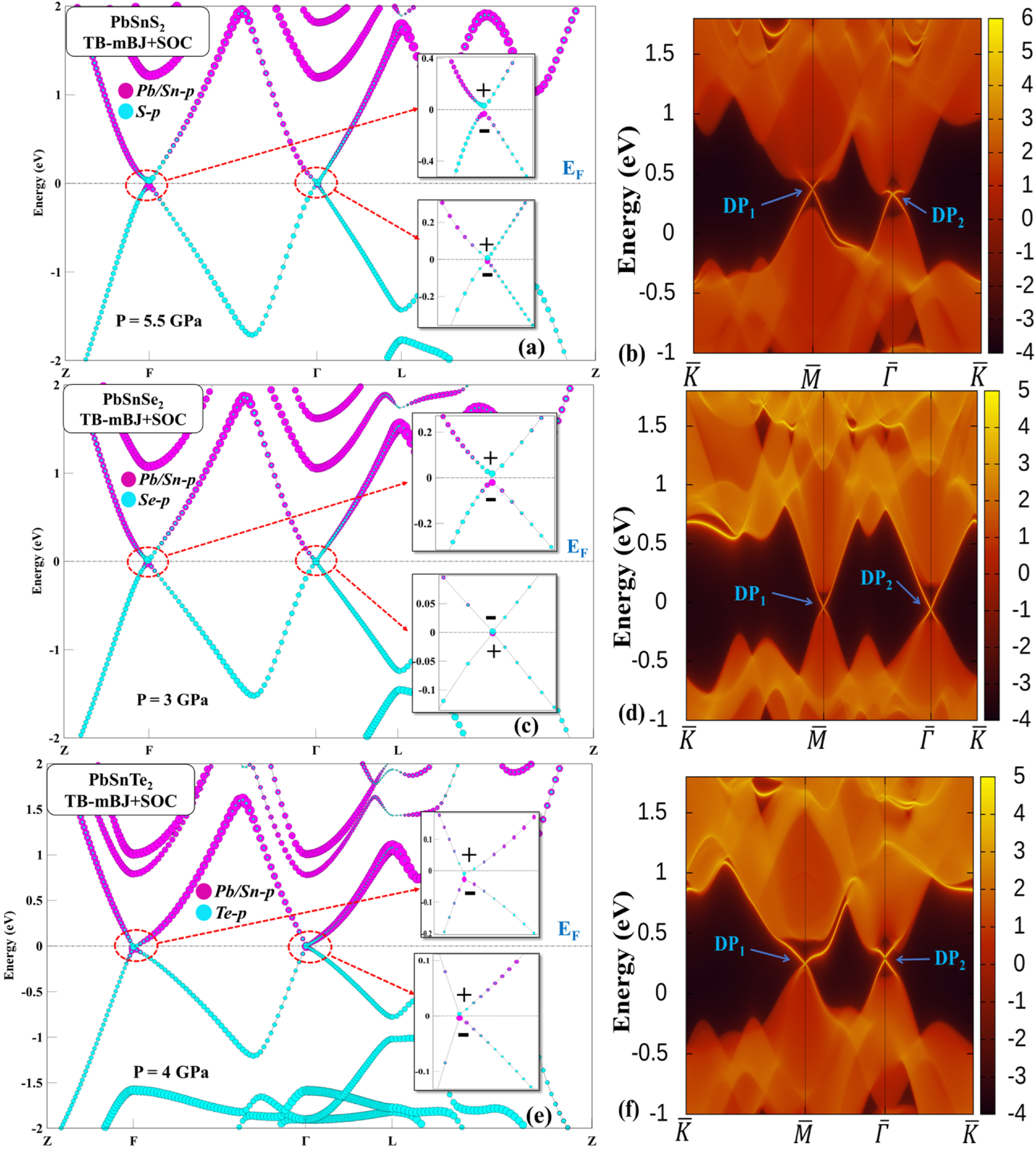}
		\caption{(a, c, e) The bulk band structure with a pair of band inversions at \textit{$\Gamma$-point} as well as \textit{F-point}, (b, d, f) The density of states (SODS) along (111) plane of the systems PbSnX$_2$ (X=S, Se, Te) at elevated pressure with marked Dirac points (DP$_1$) and (DP$_2$).} 
\end{figure}
	
	\subsection{\label{sec:level2}$\mathrm{Z_2}$ topological invariants} 
	The topologically non-trivial nature of the $\mathrm{PbSnX_2}$ (X=S, Se, Te) family under applied hydrostatic pressure can also be verified with the help of $\mathrm{Z_2}$ topological invariants. These materials have TRS as well as inversion symmetry, so their $\mathrm{Z_2}$ topological invariants can be calculated from the product of parities of all valance bands at TRIM points. According to Kane and Mele model \cite{6}, for a three-dimensional (3D) material, there are four $\mathrm{Z_2}$ topological invariants $\mathrm{(\nu_0; \nu_1\nu_2\nu_3)}$ are required to calculate at eight TRIM points, which were defined by the following equations;

	\begin{equation}
		(-1)^{\nu_0}=\prod_{n_j=0,1} \delta_{n_1 n_2 n_3}
	\end{equation}
	\begin{equation}
		(-1)^{v_{i=1,2,3}}=\prod_{n_i=1 ; n_{j \neq i}=0,1} \delta_{n_1 n_2 n_3}
	\end{equation}
	
	Here, $\delta$ and \textit{${n_i}$} are representing the parities of all the occupied bands at TRIM points and reciprocal lattice vectors, respectively. The first $\mathrm{Z_2}$ topological invariant $\mathrm{(\nu_0)}$ is independent of the choice of primitive reciprocal lattice vectors $({b_k})$ but the other three are dependent. But these other three invariants can be recognized with $G_v \equiv \sum_j v_j b_j$ which belongs to 8 element \textit{mod} 2 reciprocal lattices and can be construed as Miller indices of the reciprocal lattice vector. The calculated four $\mathrm{Z_2}$ topological invariants with the help of the product of parities at TRIM points at ambient and elevated hydrostatic pressures of the $\mathrm{PbSnX_2}$ (X=S, Se, Te) family are manifested in TABLE 2.

	\begin{table}
		\centering
		\caption{The product of parities of all valance bands at TRIM points and $\mathrm{Z_2}$ invariants under different values of hydrostatic pressure. (The detailed parity tables (TABLE S1-S3) are in supplementary information). }

		\label{tab:my_label}
		\begin{tabular}{|c|c|c|c|c|c|c|} \hline 
			Materials&  Hydrostatic Pressure (in GPa)&  \textit{$\Gamma$}&  \textit{3L}&  \textit{3F}&  \textit{Z}& ${Z_2}$-invariants\\ \hline 
			\multirow{3}{*}{$\mathrm{PbSnS_2}$} &  0&    +&    -&    +&    -&   (0; 000)\\ 
			&  5&    +&    -&    -&    -&   (1; 000)\\ 
			&  5.5&    -&    -&    -&    -&   (0; 000)\\ \hline 
			\multirow{3}{*}{$\mathrm{PbSnSe_2}$} &  0&    +&    -&    +&    -&   (0; 000)\\  
			&  2.5&    +&    -&    -&    -&   (1; 000)\\  
			&  3&    -&    -&    -&    -&   (0; 000)\\ \hline 
			\multirow{3}{*}{$\mathrm{PbSnTe_2}$}  &  0&    +&    -&    +&    -&   (0; 000)\\ 
			& 3.5&   +&   -&   -&   -&  (1; 000)\\
			& 4&   -&   -&   -&   -&  (0; 000)\\\hline
		\end{tabular}

	\end{table}
	At the ambient conditions, using equations 1 and 2, the $\mathrm{Z_2}$ topological invariants ${(\nu_0; \nu_1\nu_2\nu_3)}$ are (0;000) for the $\mathrm{PbSnX_2}$ (X=S, Se, Te) family which established the trivial semiconductor nature of this family as observed in bulk band structures (FIG. 2 (a, c, e)). When we have applied hydrostatic pressures of 5 GPa, 2.5 GPa and 3.5 GPa on $\mathrm{PbSnX_2}$ (X=S, Se, Te) materials, respectively, the first  $\mathrm{Z_2}$ topological invariant${(\nu_0}$ switched to 1 (equation 1) and using equation 2 the other three invariants ${( \nu_1\nu_2\nu_3)}$ remain (000). The non-zero value of ${(\nu_0}$ recognized these materials as strong TIs at the mentioned pressures. With further increase in pressures to 5.5 GPa, 3 GPa and 4 GPa for the $\mathrm{PbSnX_2}$ (X=S, Se, Te) family, respectively, the value of  $\mathrm{(\nu_0}$ again switched from 1 to 0 (equation 1) with existence of a pair of band inversions in the bulk band structure. Now, the systems can be topologically weak or trivial in nature depending upon the values of the remaining three invariants ${( \nu_1\nu_2\nu_3)}$ under these pressure conditions. These remaining invariants ${( \nu_1\nu_2\nu_3)}$ are (000) due to the presence of the same negative parities at all eight TRIM points using equation 2. So, these materials are topologically trivial with an even number of inversions under above mentioned elevated hydrostatic pressures. 
	
	\subsection{\label{sec:level2}Evolution of TI to TCI phase} 
	The non-zero mirror Chern number (MCN) is used to characterize the TCI phase in materials because mirror symmetry protects those Dirac cones that appear in surface electronic structures. The $\mathrm{PbSnX_2}$ (X=S, Se, Te) family has similar surface states to the TCI materials such as SnTe and PbTe \cite{37, 57, 58}, therefore, these materials can also hold TCI phase with an even number of Dirac cones in (111) plane. To confirm the TCI phase in this family, we have analyzed the  (\(\bar{1}2\bar{1}\)) surface, which has mirror symmetry around the (\(\bar{1}2\bar{1}\)) plane. The perpendicular intersection of (\(\bar{1}2\bar{1}\)) plane with (\(\bar{1}0\bar{1}\)) plane is along \(\bar{\Gamma}-\bar{X}\) k-path as shown in FIG. 1 (c). Since, $\mathrm{PbSnX_2}$ (X=S, Se, Te) materials have a pair of band inversion (FIG. 5) at $\Gamma$\textit{-} and \textit{F-points} like SnTe material, we have tried to reproduce the circumstances of SnTe in this family of materials. We anticipate the hybridization of Dirac cones at the crossing which are projected on  (\(\bar{1}2\bar{1}\)) surface. This has been confirmed in FIG. 5 (a, c, e), where the hybridization opens an energy gap along \(\bar{X}-\bar{\Gamma}\) path near the Fermi level but along the mirror symmetry path \(\bar{\Gamma}-\bar{X}\) Dirac cone holds its crossing. This Dirac cone along the \(\bar{\Gamma}-\bar{X}\) is protected by mirror symmetry (not TRS) of (\(\bar{1}0\bar{1}\)) plane. Along the \(\bar{S}-\bar{Y}-\bar{\Gamma}\) path in FIG. 5 (b, d, f), an energy gap is opened after hybridization, because the \(\bar{Y}\)\textit{-point} do not lie on the mirror plane (\(\bar{1}0\bar{1}\)), which is the projection of two \textit{F-points} \cite{42} from bulk BZ. Hence, the mirror symmetry-protected Dirac cones exist in the $\mathrm{PbSnX_2}$ (X=S, Se, Te) family, and a TI to TCI phase transition occurs after second inversion under hydrostatic pressure.
	
	\begin{figure}
		\centering
		\includegraphics[width=1\linewidth]{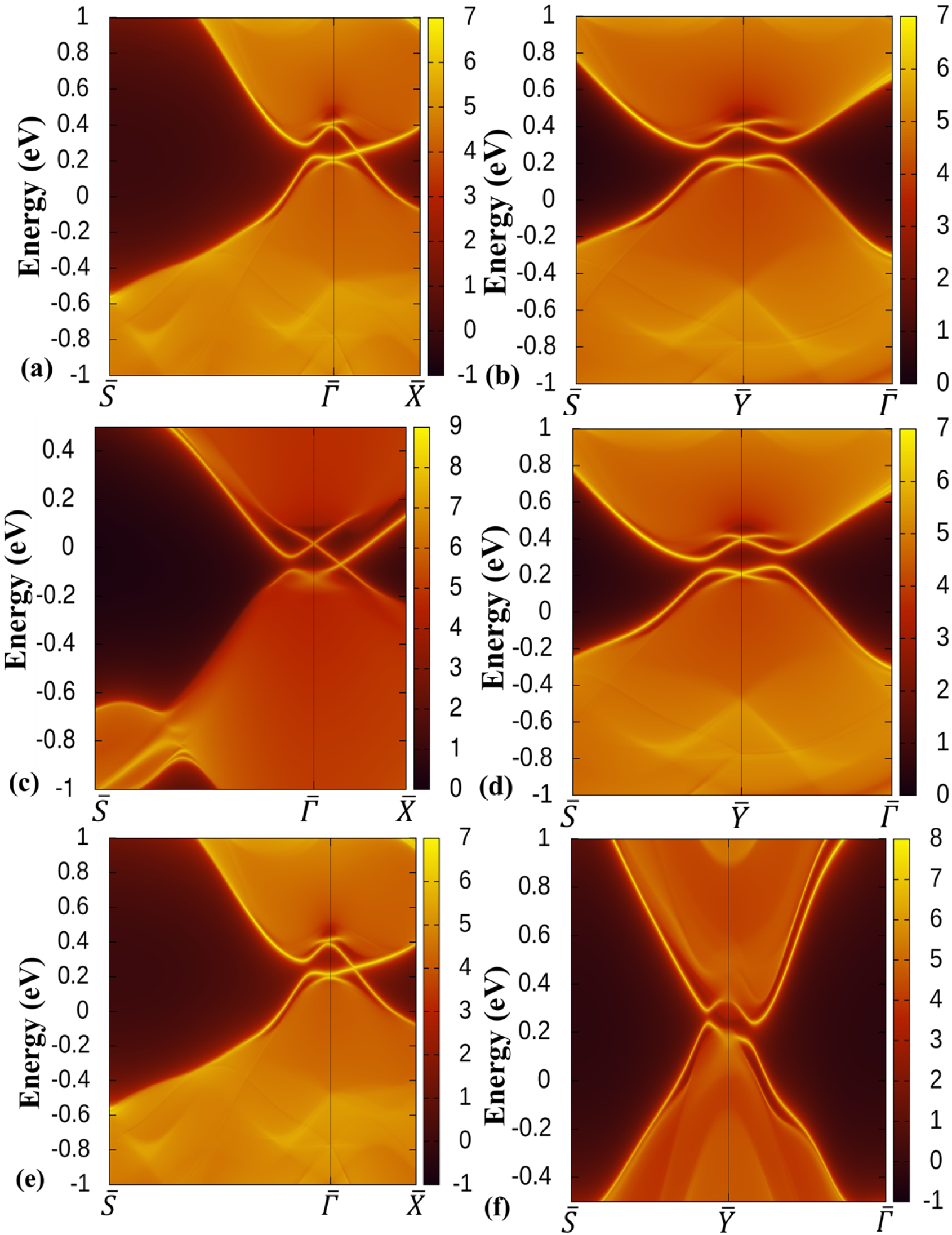}
		\caption{The SDOS of the slabs (a, b) $\mathrm{PbSnS_2}$, (c, d) $\mathrm{PbSnSe_2}$ and (e, f) $\mathrm{PbSnTe_2}$ along the \(\bar{S}-\bar{\Gamma}-\bar{X}\) and \(\bar{S}-\bar{Y}-\bar{\Gamma}\), respectively, with (\(\bar{1}2\bar{1}\)) oriented planes.}
		\label{fig:enter-label}
	\end{figure}
	This existence of TCI phase is further verified with MCN of the $\mathrm{PbSnX_2}$ (X=S, Se, Te) materials, with an even number of Dirac cones with respect to the mirror symmetry plane (\(\bar{1}0\bar{1}\)), which can be calculated using Wannier charge centers (WCC) or Wilson loop along the (111) plane. FIG. 6 (a-c) shows plots of the evolution of WCC for all occupied bands with respect to mirror operator for (\(\bar{1}0\bar{1}\)) plane. The winding of the WCC of these materials in FIG. 6 (a-c) for mirror eigen values i.e., \textit{+i} (magenta) and \textit{-i} (cyan) illustrate that MCN has value 2 along the above-mentioned mirror symmetry. The non-zero even value of MCN and mirror symmetry protected Dirac cone established the TI to TCI phase transitions in $\mathrm{PbSnX_2}$ (X=S, Se, Te) materials at 5.5 GPa, 3 GPa and 4 GPa, respectively. 
	
	\begin{figure}
		\centering
		\includegraphics[width=1\linewidth]{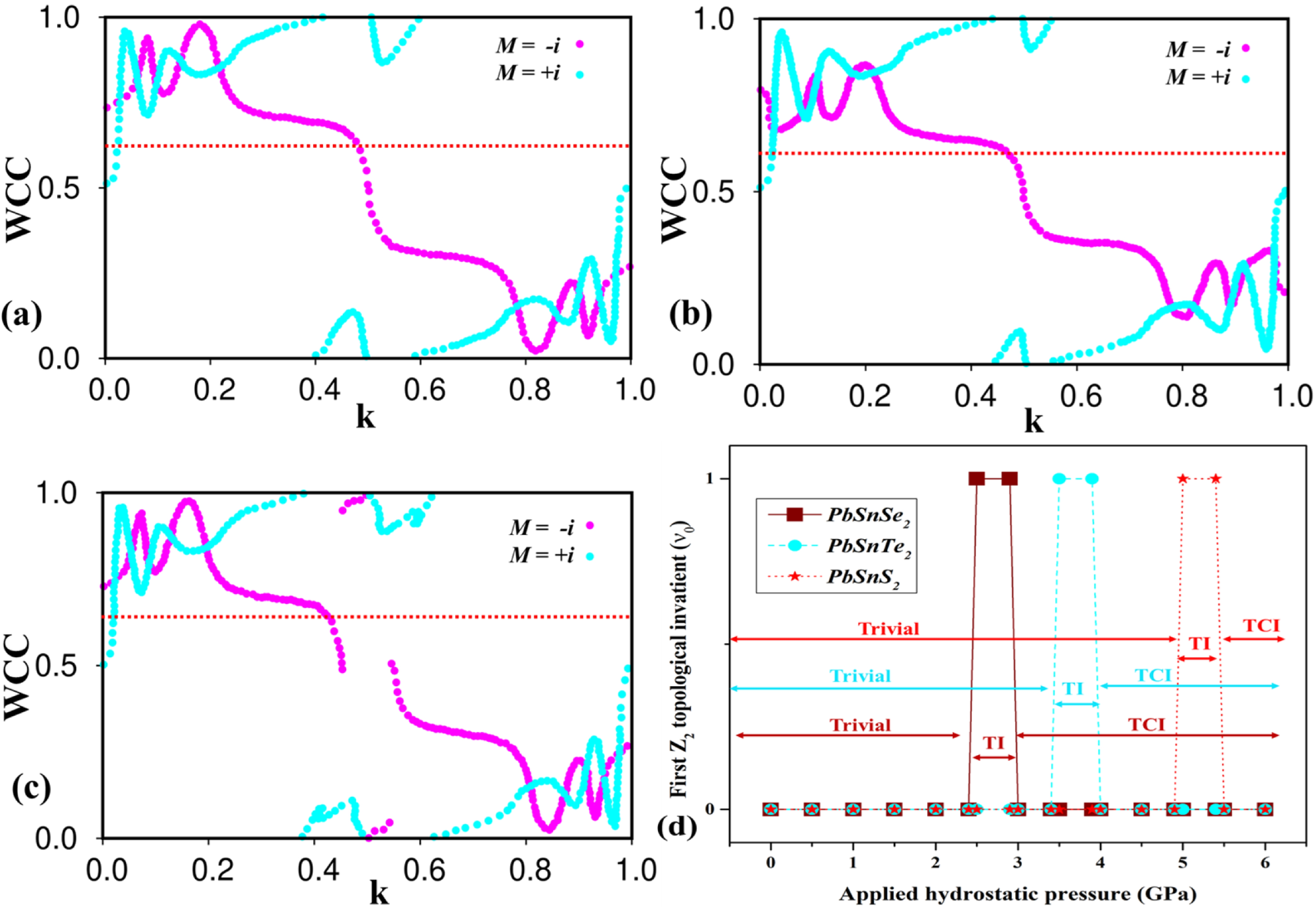}
		\caption{The evolution of the mirror eigenvalues of \textit{+i} (magenta) and \textit{-i} (cyan) in WCC for MCN in (a) $\mathrm{PbSnS_2}$, (b) $\mathrm{PbSnSe_2}$, and (c) $\mathrm{PbSnTe_2}$. (d) The variation of the first $\mathrm{Z_2}$ topological invariant $(\nu_0)$ with applied hydrostatic pressure. }
		\label{fig:enter-label}
	\end{figure}
	
	The TPT with the variation of the first $\mathrm{Z_2}$ topological invariant $\mathrm{(\nu_0)}$ with hydrostatic pressure is illustrated in FIG. 6 (d). $\mathrm{PbSnX_2}$ (X=S, Se, Te) materials have shown a transition from topologically trivial to non-trivial phase with an odd number of inversions under hydrostatic pressure of 5 GPa, 2.5 GPa and 3.5 GPa, respectively. Further increase in pressure has shown another transition from non-trivial to trivial topological phase and even number of inversions at 5.5 GPa, 3 GPa and 4 GPa, respectively.  Moreover, the symmetry analysis with  (\(\bar{1}0\bar{1}\)) mirror plane has shown the TCI phase is achieved after the second transition in these materials with an even number of Dirac cones. 
	
	\subsection{\label{sec:level2}CONCLUSION} 
	
	We have systematically studied the structural stability, and electronic and topological properties of the \textit{Sn-based} ternary chalcogenide $\mathrm{PbSnX_2}$ (X=S, Se, Te) family under ambient and applied hydrostatic pressure. The TB-mBJ functional has been identified as most relevant \textit{exchange-correlation functionals} for this study based upon the comparative analysis of the band gap of $\mathrm{TlBiS_2}$, an experientially synthesized material isostructural to the under-study materials. The phonon dispersion spectrum has confirmed the dynamical stability of these throughout the study. At ambient conditions, $\mathrm{TlBiS_2}$, $\mathrm{PbSnSe_2}$ and $\mathrm{PbSnTe_2}$ have topologically trivial semiconducting nature with direct band gap of 0.338 eV, 0.183 eV and 0.235 eV at \textit{F-point}. An increase in applied the hydrostatic pressure to 5 GPa, 2.5 GPa and 3.5 GPa, respectively, for $\mathrm{PbSnX_2}$ (X=S, Se, Te) has led to closing of energy gap at the \textit{F-point}. At these pressure values, the first TPT took place which converts these materials to TIs and the same which has been verified with an odd number of Dirac cones in SDOS and parity analysis at TRIM points. Further enhancement in applied pressure to 5.5 GPa, 3 GPa and 4 GPa, another band inversion has observed at \textit{$\Gamma$-point} in the bulk band structure with an even number of Dirac cones along the (111) plane which make these materials topological trivial again. This topologically trivial phase of these materials, however, has identified as TCI phase along (\(\bar{1}2\bar{1}\)) surface which holding mirror symmetry for (\(\bar{1}0\bar{1}\)) plane. The SDOS along the k path  \(\bar{S}-\bar{\Gamma}-\bar{X}\) and \(\bar{S}-\bar{Y}-\bar{\Gamma}\) of (\(\bar{1}2\bar{1}\)) plane has established the mirror symmetry protection for the Dirac cones. This TCI phase has further corroborated with even value of mirror Chern number calculated using the evolution of mirror eigenvalues \textit{±i} in WCC. This kind of transition from trivial to TI to TCI can also be observed in other similar ternary chalcogenides and the pressure-induced surface states can be further studied experimentally with the help of angle and spin-resolved photoemission spectroscopy.

	\begin{acknowledgments}
		All the authors acknowledge National Supercomputing Mission (NSM) for providing computing resources of ‘PARAM Siddhi-AI’, under National PARAM Supercomputing Facility (NPSF), C-DAC, Pune and supported by the Ministry of Electronics and Information Technology (MeitY) and Department of Science and Technology (DST), Government of India. One of the authors (Ramesh Kumar) would like to thank the Council of Scientific and Industrial Research (CSIR), Delhi, for financial support.
		\begin{description}
			\item[Author contributions]
			These authors contributed equally to this work.
			\item[Conflict of interest statements]
			The authors declare no competing financial interest.
		\end{description}

	\end{acknowledgments}

	%
	
\end{document}